\documentclass[conference]{IEEEtran}
\IEEEoverridecommandlockouts

\usepackage{cite}
\usepackage{mathtools}
\usepackage{enumerate}
\usepackage{braket}
\usepackage{amsmath,amssymb,amsfonts}
\usepackage{algorithmic}
\usepackage{graphicx}
\usepackage{textcomp}
\usepackage{hyperref}
\usepackage{subfig}
\usepackage{booktabs}  
\usepackage{longtable}
\usepackage{adjustbox}
\usepackage[table,xcdraw]{xcolor}

\usepackage{soul} 

\begin{document}
\title{Crosstalk-Based Parameterized Quantum Circuit Approximation\thanks{This work was supported in part by the QISE-NET NSF Fellowship DMR 17-47426}}
\author{\IEEEauthorblockN{Mohannad Ibrahim\IEEEauthorrefmark{1}, Nicholas T. Bronn\IEEEauthorrefmark{2}, and Gregory T. Byrd\IEEEauthorrefmark{1}}
\IEEEauthorblockA{\IEEEauthorrefmark{1}\textit{Dept. of Electrical and Computer Engineering, NC State University, Raleigh, North Carolina, USA} \\
\IEEEauthorrefmark{2}\textit{IBM Quantum, IBM T.J. Watson Research Center, Yorktown Heights, New York, USA} \\
mmibrah2@ncsu.edu\IEEEauthorrefmark{1},
ntbronn@us.ibm.com\IEEEauthorrefmark{2}
gbyrd@ncsu.edu\IEEEauthorrefmark{1}}
}
\maketitle

\begin{abstract}

In this paper, we propose an ansatz approximation approach for variational quantum algorithms (VQAs) that uses one of the hardware's main attributes, its crosstalk behavior, as its main approximation driver. By utilizing crosstalk-adaptive scheduling, we are able to apply a circuit-level approximation/optimization to our ansatz. Our design procedure involves first characterizing the hardware's crosstalk and then approximating the circuit by a desired level of crosstalk mitigation, all while effectively reducing its duration and gate counts. We demonstrate the effect of crosstalk mitigation on expressibility, trainability, and entanglement: key components that drive the utility of parameterized circuits. We tested our approach on real quantum hardware against a base configuration, and our results showed superior performance for the circuit-level optimized ansatz over a base ansatz for two quantum chemistry benchmarks. We take into consideration that applications vary in their response to crosstalk, and we believe that this approximation strategy can be used to create ansatze that are expressive, trainable, and with crosstalk mitigation levels tailored for specific workloads.
\end{abstract}


\section{Introduction}
Near-term quantum computers are characterized by a limited number of qubits, in the range of 10s to 100s with current technology. Because there are not enough qubits to implement full-scale quantum error correction, system and environmental noise is exposed to the algorithm, which limits the useful depth of quantum circuits. Variational quantum algorithms are a promising approach for current hardware, utilizing reasonably short-depth circuits and tunable parameters that can help mitigate systemic noise.

A \textit{variational quantum algorithm (VQA)} is a hybrid scheme of computation that allocates tasks to both quantum and classical computing resources and coordinates the execution between the two through a tight feedback loop. The quantum computer's task is to prepare and measure relevant quantum states generated by the so-called \textit{ansatz} or \textit{Parameterized Quantum Circuit (PQC)}. The classical computer's task is to update/optimize the circuit parameters, which are then fed back into the quantum computer to prepare a new state. This cycle is repeated until some convergence criteria are satisfied.

VQAs have been applied to a wide variety of applications \cite{vqas} such as quantum chemistry \cite{kandala, chem_1,chem_2,chem_3}, combinatorial optimization \cite{comb_1,comb_2,comb_3}, and machine learning \cite{expr_5,entang_3, ml_1, ml_2, expr_4}. A widely-used VQA is the\textit{ Variational Quantum Eigensolver (VQE)} \cite{vqe}, which seeks to find the minimum eigenvalue of a matrix. When used in quantum simulation, the matrix is typically the Hamiltonian of a system. However, the algorithm is not just limited to finding low energy eigenstates; it can be extended to minimize any objective function that is expressible as a quantum observable~\cite{cerezo_bp, qeom, Higgott_2019}.

In this paper, we propose a different approach for VQA optimization by integrating one of the quantum hardware's characteristics, particularly its crosstalk noise, in the design process of PQCs. We propose an approximation strategy that uses the hardware's crosstalk behavior to create approximate versions of a PQC with different levels of crosstalk mitigation. We refer to PQCs created using this technique as Crosstalk-optimized (Xtalk) PQCs. Multiple hardware features, such as the native gate set, topology, and noise model, are unique to each machine. We chose crosstalk as our main hardware characteristic as it is recognized as a major challenge in quantum computing, mitigable by both hardware and software techniques \cite{sheldon_cr, nereeja_cr, murali_xtalk, ding_xtalk}. Additionally, crosstalk-adaptive scheduling techniques \cite{murali_xtalk, ding_xtalk} make it possible for this attribute to be easily integrated into PQC design, as we will further discuss in later sections.

Crosstalk can be defined as a mixture of unwanted interactions between coupled qubits in a quantum device~\cite{ding_xtalk, murali_xtalk}. It can appear in many architectures such as trapped ions~\cite{t_ions} and superconducting systems~\cite{sheldon_cr, nereeja_cr, krantz_guide}. Crosstalk errors can manifest in many ways: as an exchange of excitation, leakage to non-computational states \cite{xtalk_diabatic}, or an order of magnitude worse gate fidelities \cite{murali_xtalk}, which are all detrimental to quantum programs.

In superconducting systems, crosstalk can occur for multiple reasons. IBM superconducting devices, for example, use fixed-frequency resonators to couple their fixed-frequency transmon qubits. This coupling produces an always-on $ZZ$ interaction proportional to the coupling strength or transmon-transmon exchange $J$ \cite{zz, sheldon_cr}. This always-on $ZZ$ interaction is a major source of error in fixed-frequency devices. Besides reducing the two-qubit gate fidelities below the levels set by coherence, interactions with spectator qubits (qubits that are not part of the two-qubit interaction) cause unwanted entanglement to accumulate across the system \cite{nereeja_cr}.

Crosstalk mitigation is, therefore, a major goal in quantum hardware design and fabrication. For example, IBM machines employ a set of crosstalk mitigation techniques, such as limiting the device's connectivity---which effectively simplifies frequency allocation and limits the number of spectators to a two-qubit interaction while admitting a quantum error correction code --- and laser annealing of Josephson junctions~\cite{chamberland2020, Hertzberg2021}. Fixed-frequency transmon architectures such as the one proposed in~\cite{Stehlik2021} utilize tunable couplers to achieve faster two-qubit gates and to address errors due to the always-on $ZZ$ term. Other tunable superconducting devices, such as Google's Sycamore processor~\cite{google_qec}, where both qubits and couplers are tunable, scalable optimization techniques are utilized to extract error-reducing frequencies~\cite{snake}. On a pulse level, experimental results in~\cite{nereeja_cr} proved that adding rotary echoes to the two-qubit cross-resonance interaction suppresses errors arising from the static $ZZ$ term. Despite such efforts, crosstalk still exists in today's quantum machines, as we will see in Section~\ref{sec:cr_char}, and is still one of the main scalability-related challenges.

Leveraging crosstalk-aware scheduling, we develop a novel ansatz approximation mechanism that is capable of creating different versions of a base configuration with different levels of crosstalk mitigation, described in Section~\ref{sec:xtalk-pqc-approximation}. In Section~\ref{sec:eval}, we evaluate the effectiveness of the approach. Our crosstalk-based ansatze outperform the base configuration for two quantum chemistry benchmarks, speeding up execution by up to $2.9\times$ and $1.83\times$ on average. Moreoever, we explore the connection between crosstalk and trainability by demonstrating that circuits experiencing more crosstalk have lower trainability.


\section{Background}
\subsection{Randomized Benchmarking}
\label{sec:rb}
Characterizing the noise affecting a quantum system is useful in many ways. It allows for many optimizations to a workload's execution on the system and for good error-correction schemes. Randomized Benchmarking (RB) is a widely-used, scalable technique for partially characterizing a quantum system's noise. It is used by quantum hardware developers to benchmark known gate sets such as Clifford and CNOT-Dihedral~\cite{rb_1, rb_2, rb_3, stancil_byrd}. 
The RB protocol can be summarized in four steps~\cite{rb_2}:
\begin{enumerate}[\bfseries Step 1:]
\item $K$ sequences of different lengths ($m$) are generated. Each sequence consists of random gates from a specific gate set (e.g., Clifford) and a computed inverse to return the qubits to their initial state.
\item The sequences are executed on the hardware under investigation. Each sequence is modeled for later processing with a variable $S_\mathrm{i_m}$ that accounts for the error rate of each operation in the sequence.
\item The survival probability $\mathrm{Tr}[E_\psi S_\mathrm{i_m}(\rho_\psi)]$ of each of the $K$ sequences is measured, where $\rho_\psi$ is the initial state (taking into account initial state preparation errors), and $E_\psi$ is the positive operator-valued measurement (POVM) that takes into account measurement errors. The average fidelity for the sequences $K_m$ is then calculated
\begin{equation}
F_\mathrm{seq} = (m,\ket{\psi}) = Tr[E_\psi S_\mathrm{K_m}(\rho_\psi)]
\end{equation}
where the average sequence operation $S_\mathrm{K_m} = \frac{1}{K_m} \sum_\mathrm{i_m} S_\mathrm{i_m}$.
\item The experiment is repeated for different sequence lengths ($m$). The average sequence fidelities obtained in Step 3 are fitted to
\begin{equation}
F^{(0)}_\mathrm{seq} = (m, \ket{\psi}) = A_0 \alpha^m + B_0 
\end{equation}
where $F^{(0)}$ is the gate-independent and time-independent ``simpler'' fitting model. $A_0$ and $B_0$ coefficients encode the state preparation and measurement errors, respectively. The average error rate or Error per Clifford (EPC) is determined by the parameter $\alpha$ through
\begin{equation}
\mathrm{EPC} = 1-\alpha - \frac{1-\alpha}{2^n}
\end{equation}
where $n$ is the number of the qubits.
\end{enumerate}
\begin{figure}[!t]
    \centering
    \includegraphics[width=0.8\linewidth, keepaspectratio]{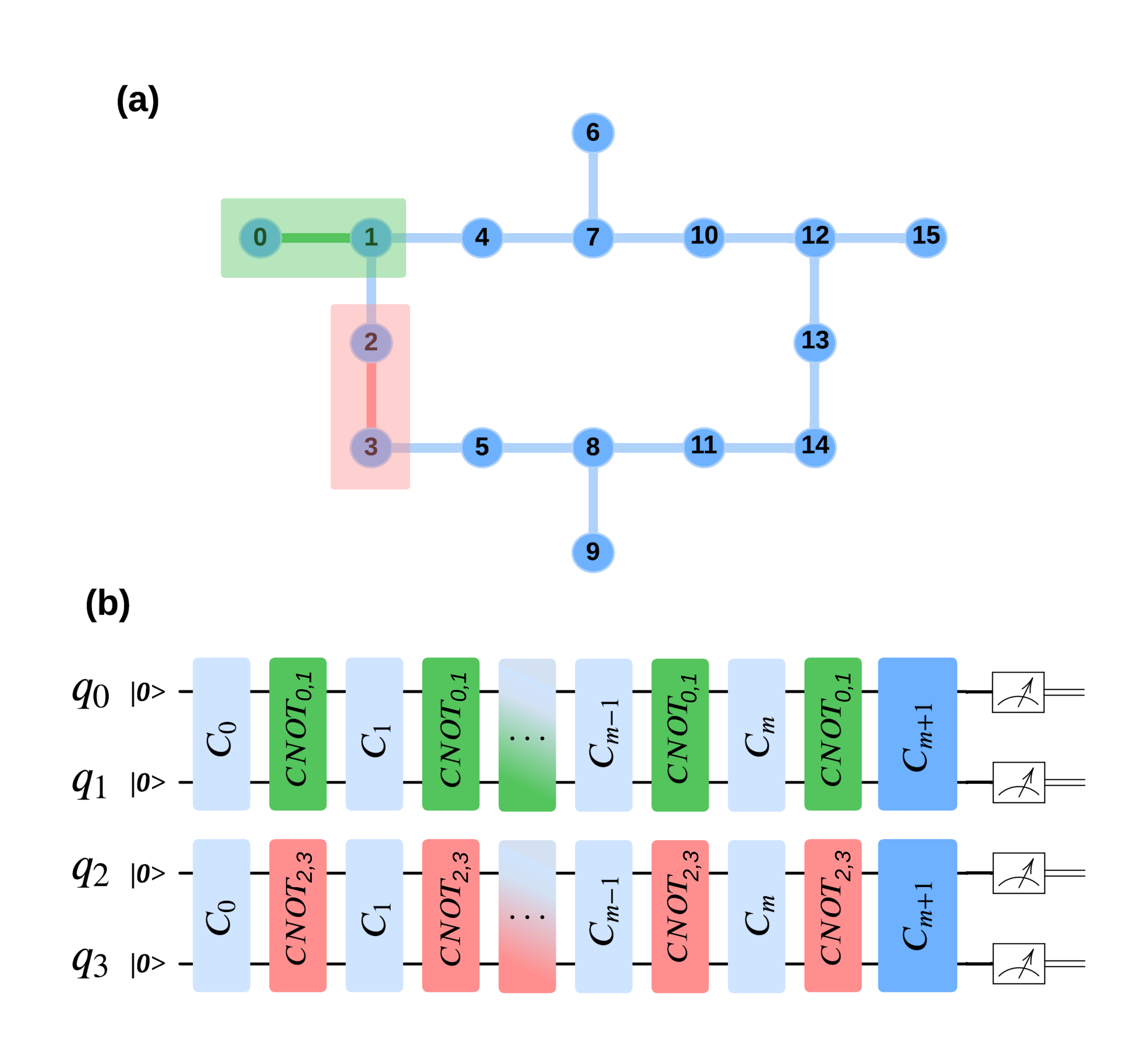}
\caption{\textbf{(a)} The coupling map of \textit{ibmq\_guadalupe}. \textbf{(b)} Simultaneous Randomized Benchmarking (SRB) for CNOT$_{0,1}$ and CNOT$_{2,3}$ with $\{C_0,...,C_m\}$ being the random Cliffords and $C_{m+1}$ the inverting Clifford. The mapping of the gates on the backend is indicated by the green and red highlighting in (a). Operations indicated by the colored regions can be run in parallel as they do not share resources. Note also that the coupling between qubits $1$ and $2$ cannot be used when both qubits are busy.}
    \label{fig:guadalupe_and_srb}
\end{figure}
\textit{Simultaneous} RB (SRB)~\cite{srb}, which consists of RB experiments run simultaneously on sets of qubits, allows for further investigations of a system's noise properties by comparing to RB experiments run individually. It enables the measurement of crosstalk and ``conditional'' error rates: gate errors on a qubit while nearby qubits are active. Utilizing SRB to control and handle errors arising from crosstalk and unwanted interactions in multi-qubit systems has been proposed in~\cite{murali_xtalk}. We further explain their proposed methodology in the next section.
\begin{figure}[t]
    \centering
    \includegraphics[width=\linewidth, keepaspectratio]{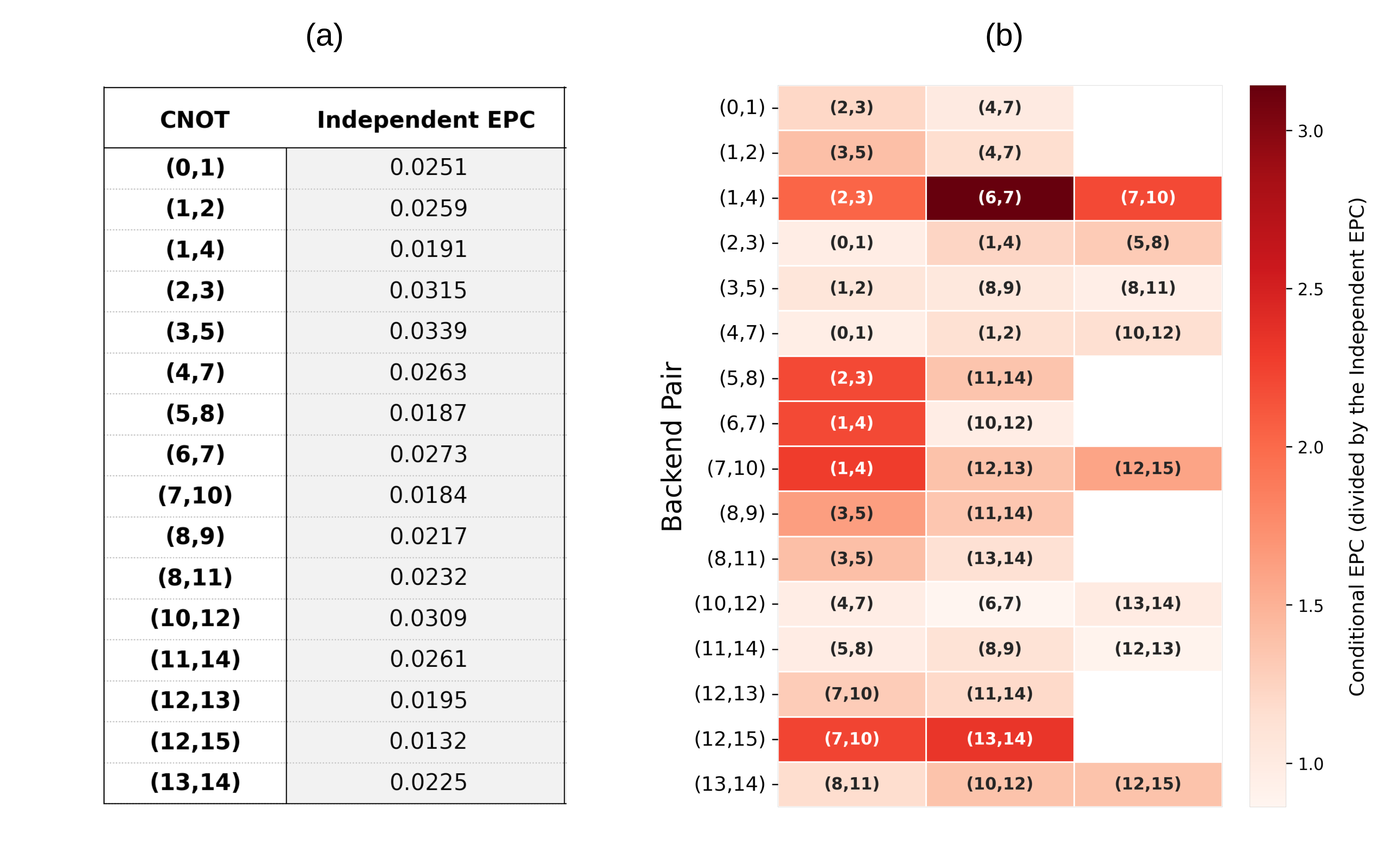}
    \caption{\textbf{(a)} Individual Error-per-Clifford (EPC) rates for CNOTs executed on \textit{ibmq\_guadalupe}. \textbf{(b)} Conditional Error-per-Clifford (EPC) rates when two CNOTS are executed simultaneously. Cells are shaded according to the relative conditional-to-independent error rates.}
    \label{fig:xtalk_data}
\end{figure}
\subsection{Expressibility, Trainability \& Entanglement}
To evaluate whether a PQC can prepare the target quantum state, different metrics have been proposed \cite{sukin_expr, expr_2, expr_4, expr_5, holmes_expr}. In this section, we describe three qualitative metrics used in this paper to estimate a PQC's \textit{expressibility}, \textit{trainability}, and \textit{entanglement}.

\textbf{Expressibility} is a metric first proposed by Sim \textit{et al.}~\cite{sukin_expr} to evaluate a PQC's ability to produce quantum states that closely represent the Hilbert space. This is done by comparing the distribution of states obtained from a PQC's parameterized unitary to the maximally expressive uniform \textit{(Haar)} random states. It is estimated using the Kullback-Leibler (KL)~\cite{kl} divergence as follows
\begin{equation}
{\rm Expr} = D_{\text{KL}|}(\hat{P}_{\text{PQC}}(F;\vec{\theta}) || P_{\text{ Haar}}(F))
\end{equation}
where $P_{\text{ Haar}}(F)$ is the probability distribution of fidelities $F$ for the Haar random state and $\hat{P}_{\text{PQC}}(F;\vec{\theta})$ is the probability distribution of the fidelities of quantum states prepared by the PQC. As a divergence measure, a smaller expressibility value indicates a more expressive circuit.

\textbf{Trainability}: the trainability of PQCs is crucial for achieving good performance in VQAs. However, simply increasing the expressiveness of a PQC does not always lead to better performance. It is crucial to characterize the optimization landscape of a VQA and use efficient training routines to ensure good performance. Interestingly, perfectly expressive ansatze often have flatter optimization landscapes and are less trainable~\cite{holmes_expr}. The trainability of PQCs, particularly hardware-efficient ones, has been studied extensively, with earlier work by McClean \textit{et al.}~\cite{mcclean_bp} demonstrating the phenomenon of barren plateaus, where gradients vanish exponentially as the number of qubits increases. This observation has been further investigated by Cerezo \textit{et al.}~\cite{cerezo_bp} to show that the occurrence of barren plateaus is cost-function-dependent for shallow ansatze. Other factors, such as noise and entanglement, can also impact barren plateaus~\cite{noise_bp, ent_bp, patti_bp}. 

In Section~\ref{sec:expr_res}, we evaluate our PQCs' trainability using cost-function-dependent barren plateau analysis. Specifically, we calculate $Var[\partial_iC]$, which represents the variance of the partial derivative of a cost function $C$ with respect to parameter $\theta_i$ for $n$ sampled circuits. The magnitude of the variance reflects the concentration of the partial derivative around zero, with smaller values indicating less trainability for the PQC.

\textbf{Entanglement} measurement quantifies the amount of entanglement contained in a quantum state. Highly-entangled PQCs can capture non-trivial correlations in quantum data and efficiently represent solution spaces for tasks like ground state preparation or data classification~\cite{sukin_expr, entang_1, kandala, entang_2, entang_3}. However, excessive entanglement can lead to concentration of measure, making PQCs too random and less trainable. In recent works, entanglement has been investigated as a primary source of barren plateaus and its tradeoffs with trainability vary across optimization problems~\cite{ent_bp, patti_bp}. Thus, a comprehensive understanding of the role of entanglement in VQAs is important \cite{how_much_ent, ent_extra_1, ent_extra_2, ent_extra_3}.

In this paper, we use the bipartite entanglement entropy, which is the Von Neumann entropy of the reduced density matrix of any of the subsystems, to estimate the spread $S$ of circuit entanglement
\begin{equation}
\label{eqn:entropy}
S = Tr[\rho_\alpha\log_2\rho_\alpha]
\end{equation}
where $\rho_\alpha$  is the reduced density matrix of $(n-1)/2$ connected qubits containing as many cost function qubits as possible \cite{patti_bp}.
\section{Crosstalk Characterization using SRB}
\label{sec:cr_char}
Following the method proposed by Murali \textit{et al.}~\cite{murali_xtalk}, we characterize quantum devices' crosstalk using SRB. We focus on the effect of crosstalk on simultaneous two-qubit gates only in this paper, as we are interested in approximating the entangling layers of a PQC, which are normally comprised of two-qubit gates only. Note that this is different from crosstalk characterization at the two-qubit Hamiltonian level, achieved through tomography-based techniques~\cite{nereeja_cr, mond_hpqc}. Here, we are examining the quantum device's crosstalk behavior at a gate level.

Fig.~\ref{fig:guadalupe_and_srb}(a) shows the coupling map of IBM's $16$-qubit device \textit{ibmq\_guadalupe}, which we use to demonstrate our technique. We utilize Qiskit Experiment's~\cite{qiskit_experiments} RB infrastructure for our experiments. First, we perform \textit{Interleaved} RB (IRB) to measure the independent error rates of CNOT gates applied on each backend pair, with no operations executed in parallel. IRB interleaves the gate under investigation (CNOT) with multiple random sequences generated from the gate set~\cite{rb_1, rb_2, rb_3}. As we discussed in Section~\ref{sec:rb}, the measured results are fitted to a theoretical model that accounts for the measured qubits' ground state population, from which a CNOT's error is estimated. The results from this experiment are shown in Fig.~\ref{fig:xtalk_data}(a).

Next, we perform SRB by performing IRB on two CNOTs simultaneously, as shown in Fig.~\ref{fig:guadalupe_and_srb}(b). Two operations can be parallelized if they do not share a quantum resource (both qubits and couplers in this case), as shown by the colored links in Fig.~\ref{fig:guadalupe_and_srb}(a). SRB allows us to measure the conditional error rate of CNOT$_{0,1}$ in the presence of CNOT$_{2,3}$ and vice versa. When the conditional error rate is higher than the independent rate, we generally attribute that difference to crosstalk interference. Fig.~\ref{fig:xtalk_data} illustrates the \textit{independent} and \textit{conditional} error rates for CNOT gates executed on \textit{ibmq\_guadalupe}. The experiment reveals that multiple parallel CNOTs incur crosstalk at different levels of severity that can degrade their error rate by up to $3.14\times$. 
The number of experiments is reduced by performing SRB on CNOT pairs that are only one hop away from each other on the coupling map. Experimental results from~\cite{murali_xtalk} prove that crosstalk noise is significant only at this distance for IBM machines, which comes as a natural result of the device's limited connectivity. Additionally, results from \cite{ash_xtalk} demonstrated that simultaneous single- and two-qubit gates SRB show minor changes in error rates due to crosstalk. However, the method is still applicable if such a level of characterization is desired.

\section{Crosstalk-Based PQC Approximation}
\label{sec:xtalk-pqc-approximation}
In this section, we describe our PQC approximation approach in detail. Our goal is to integrate crosstalk in the design process of PQCs. Such integration allows for further understanding and evaluation of the effects of noise in general, and crosstalk in specific, on the performance of VQAs, and characteristics of PQCs such as \textit{expressibility}~\cite{sukin_expr} and \textit{trainability}~\cite{mcclean_bp, cerezo_bp}. In Section~\ref{sec:eval}, we demonstrate how crosstalk closely affects all of these aspects.

Our approach is aimed at Hardware Efficient Anstaze (HEA), as they do not typically encode any problem-specific data in their structure, allowing for more approximation and rearrangement flexibility.

\subsection{Approximation to Alternating Layered Ansatz}
\label{sec:ala_approx}
\begin{figure}[!t]
    \centering
    \includegraphics[width=0.9\linewidth, keepaspectratio]{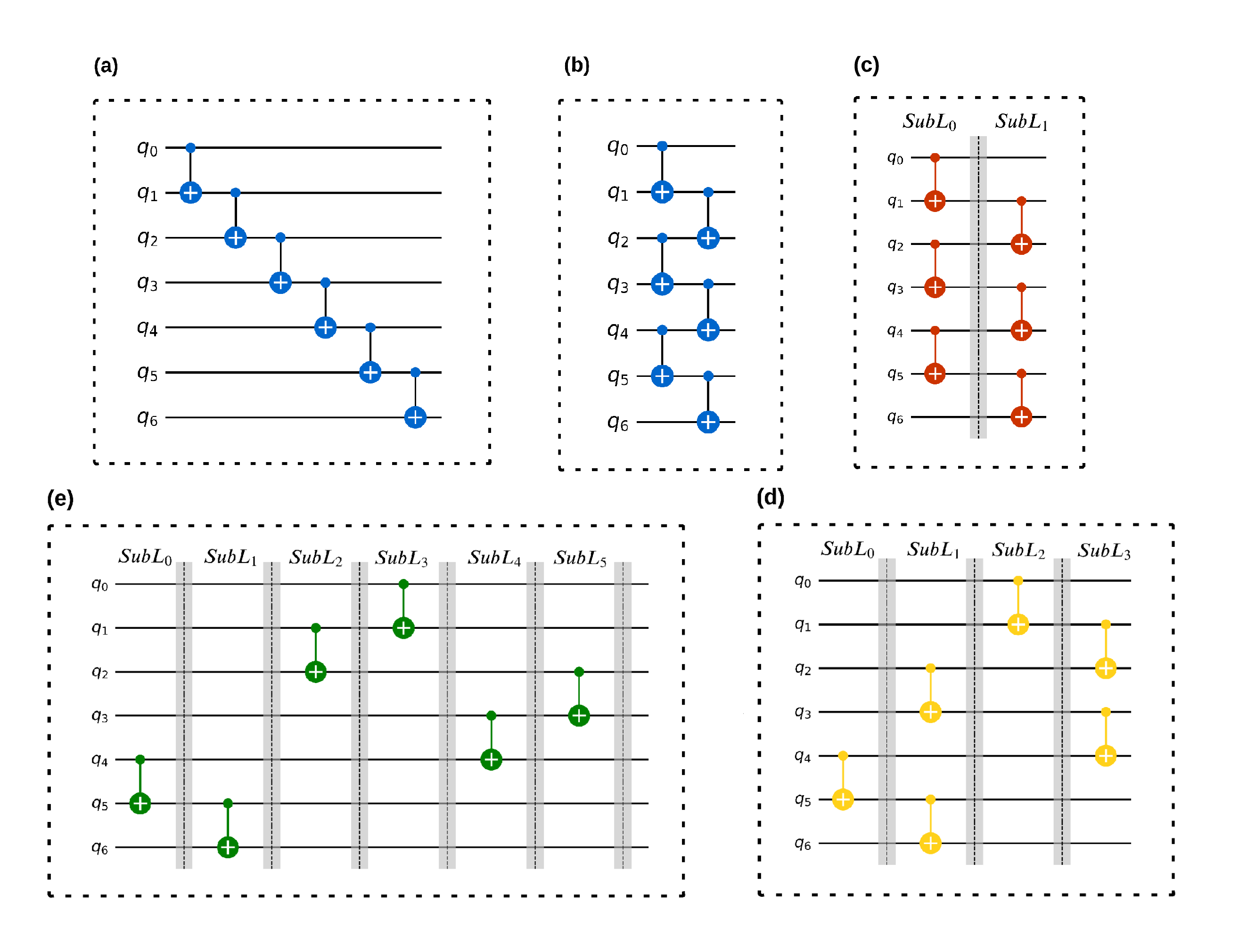}
    \caption{\textbf{(a)} The base PQC. \textbf{(b)} The PQC after applying the approximation step described in Section~\ref{sec:ala_approx}. The PQCs \textbf{(c)},  \textbf{(d)}, and \textbf{(e)} show the approximated PQC after applying Xtalk scheduling with \textbf{high}, \textbf{medium}, and \textbf{low} crosstalk tolerance respectively.}
    \label{fig:xtalk_approx}
\end{figure}

The first step in this approach is to break any ordering or dependency constraints in the PQC's entangling layer. This is achieved in Qiskit's transpiler~\cite{qiskit} by parsing the PQC's Directed Acyclic Graph (DAG) representation and identifying all operations in its first entangling layer (or sub-layer). Next, the operations are scheduled on the backend's qubits (mapped) with the maximum allowed parallelism, disregarding their ordering or commutativity constraints, resulting in a transformed DAG. In other words, this step approximates the PQC to a structure similar to an Alternating Layered Ansatz (ALA) \cite{mcclean_bp, cerezo_bp, keio_ala}. Fig.~\ref{fig:xtalk_approx}(a) and (b) show an example of a PQC configuration before and after applying this approximation step.

\subsection{Overview of Crosstalk-Adaptive Scheduling}
In this section, we give an overview of \textit{XtalkSched} \cite{murali_xtalk}, a crosstalk-adaptive scheduling algorithm that aims to mitigate the impacts of crosstalk and decoherence on a quantum program simultaneously. We employ this algorithm to extract crosstalk-mitigated sub-layers from the approximated PQC we obtained in the previous step.
XtalkSched models scheduling of the quantum circuit as a constrained optimization problem and solves it using Satisfiability Modulo Theory (SMT)~\cite{smt}. Its cost function incorporates crosstalk data, program dependencies, and machine calibration (independent error rates, coherence time, and gate duration).

To model crosstalk error, the optimizer connects the independent gate error rates with the different overlap scenarios that can happen when multiple gates are executed simultaneously. It does so by creating an overlap set for each gate \textit{Olap}($g_i$) that tracks all gates that can possibly overlap with it. Each gate pair is then assigned an overlap indicator $\sigma_{ij}$ that is set to $1$ when gates $g_i$ and $g_j$ overlap and $0$ otherwise. These overlap indicators are used to formulate the gate error constraints. For example, consider a scenario where \textit{Olap}($g_1$)$=\{g_2, g_3\}$ which creates four possible scheduling scenarios (See \cite[eq. (3)--(6)]{murali_xtalk}). As such, the overlapping scenario will determine whether the optimizer picks an independent error rate for $g_1$ (i.e., $E(g_1)$) or a conditional error rate (e.g., $E(g_1|g_2)$), which represents the error rate of $g_1$ in the presence of $g_2$. For any overlap scenarios, the scheduler is configured to pick the maximum error rate possible for a gate.

Qubit decoherence errors are accounted for by computing the \textit{lifetime} of each qubit in the schedule $q_i.t$, which is the difference between the start time of the first operation and the finish time of the last one executed on $q_i$. If a program performs a computation for time $t$ on a qubit, the probability of error from $T_1$ and $T_2$ losses are $(1-e^{\frac{-t}{T_1}})$ and $(1-e^{\frac{-t}{T_2}})$ respectively.  With that, the decoherence error is calculated as
\begin{equation}
\label{eqn:coherence}
    q_i.\epsilon = 1-e^{\frac{q_i.t}{q_i.T}}
\end{equation}
where $T$ is the minimum of $T_1$ and $T_2$, corresponding to the maximum available compute time on $q_i$. 

The optimizer also adds constraints to satisfy data dependencies as well by ensuring that if two gates $g_i$ and $g_j$ operate on the same set of qubits, the program order is satisfied. Finally, the overall cost function can be represented as follows
\begin{equation}
    \min \ \left(\omega \underbrace{\sum_{\forall g \in G}(\log \ g.\epsilon)}_{\text{Gate errors (Crosstalk)}} \ - \ (1-\omega) \underbrace{\sum_{\forall q \in Q}(q.t/q.T)}_{\text{Decoherence errors}}\right)
\label{eqn:xtalk_final}
\end{equation}
where the first term aims to minimize the gate error $g.\epsilon$ of each gate from the program gates $G$. The second term minimizes the decoherence error of each of the program qubits $Q$ according to~(\ref{eqn:coherence}). Finally, $\omega \in [0, 1]$ is a user-set parameter controlling the weight (importance) of each term, which can be tuned per application to balance between gate errors and decoherence and achieve better results.
\begin{figure}[t]
    \centering
    \includegraphics[width=\linewidth, keepaspectratio]{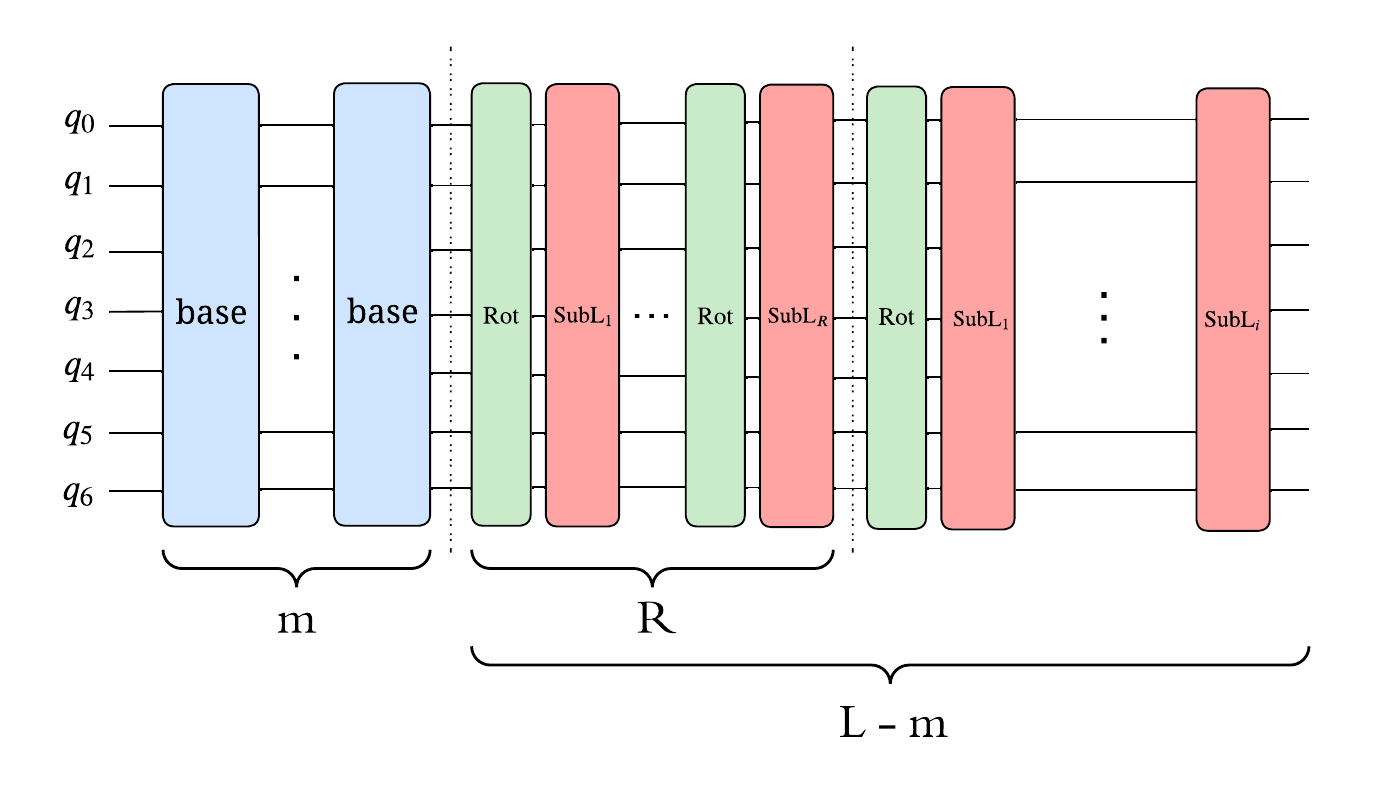}
    \caption{The \textit{Xtalk\_pqc} configuration. $m$ specifies the number of base layers. $R$ is the number of sub-layers obtained by the scheduler and is determined by the configured crosstalk mitigation level. High crosstalk mitigation will lead to larger $R$ and vice-versa.}
    \label{fig:clo_pqc}
\end{figure}
\subsection{Alternating Crosstalk-Mitigated Layers}
\label{sec:clo_method}
In this step, we apply XtalkSched to the approximated circuit obtained in the first step (Section~\ref{sec:ala_approx}) to extract ``crosstalk-mitigated'' sub-layers that we can use in our Xtalk PQCs. To enable the scheduler to accurately provide us with different levels of crosstalk mitigation, we modify two parameters:
\begin{itemize}
\item First, we increase the threshold used to calculate each gate's \textit{Olap} set. This ensures that all conditional error rates larger than $1$ are accounted for, essentially making the scheduler more sensitive to crosstalk.
\item With all possible overlaps now considered, alternating the value of $\omega$ in~(\ref{eqn:xtalk_final}) gives us the desired outcome of different levels of crosstalk mitigation. Dropping $\omega$ to $0$ forces the scheduler to create sub-layers with maximum parallelism, as the cost function will only optimize for decoherence. Increasing $\omega$ balances between the two error terms up until $1$, at which the scheduler optimizes for gate errors only (including crosstalk) and completely serializes the execution.
\end{itemize}
Once the schedules are obtained, the scheduler adds controls in the form of \textit{barriers} as a post-processing step. This is important to our approach as it facilitates extracting the sub-layers from the scheduled IR. Figures~\ref{fig:xtalk_approx}(c), (d), and (e) show the result of scheduling the approximated circuit with three levels of crosstalk tolerance: \textit{high} ($\omega=0$), \textit{medium} ($\omega=0.5$), and \textit{low} ($\omega=1$) respectively.
\begin{figure}[!t]
    \centering
    \includegraphics[width=0.8\linewidth, keepaspectratio]{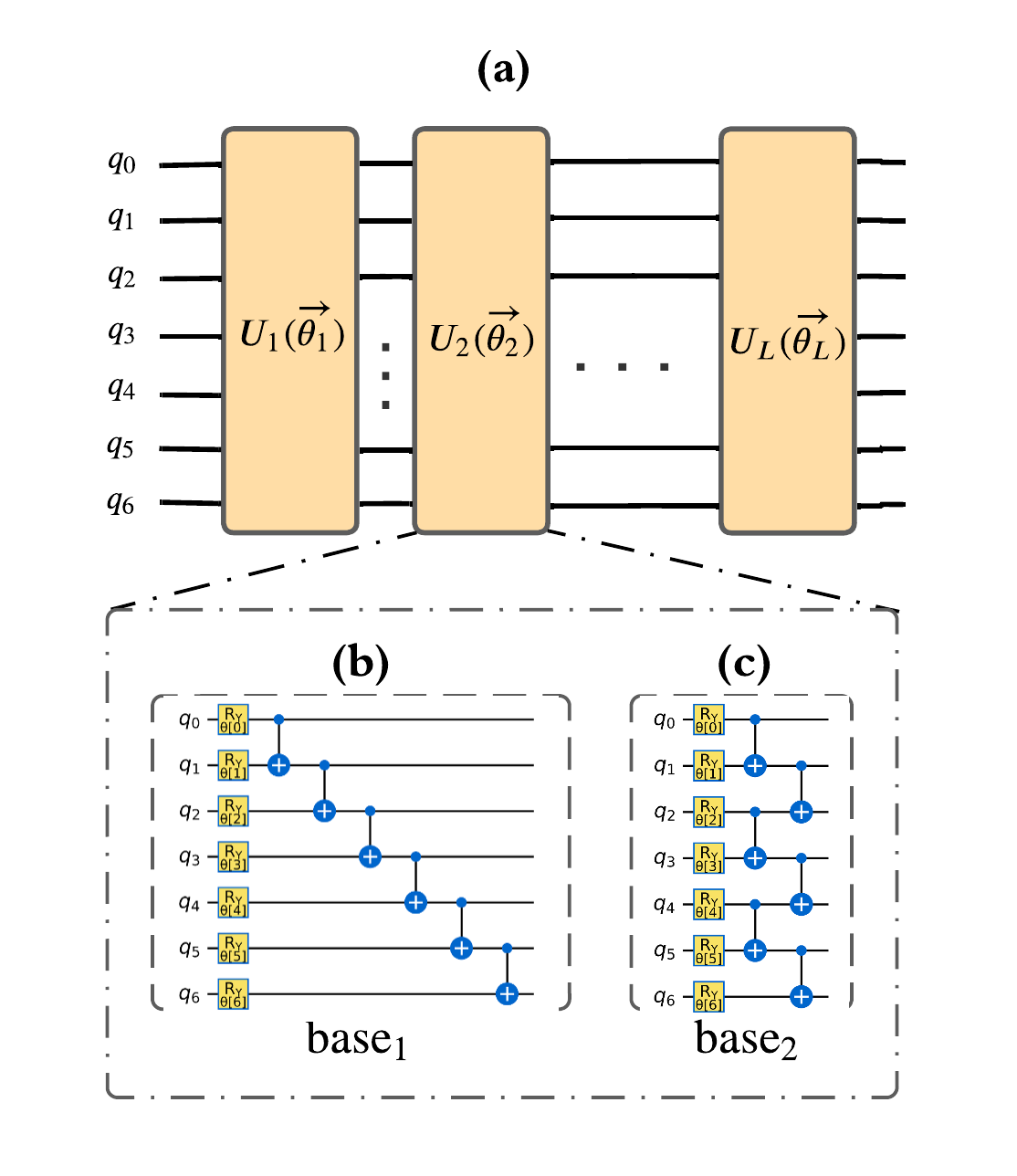}
    \caption{\textbf{(a)} Base PQC configuration. \textbf{(b)} and \textbf{(c)} show the layer configuration for \textit{base}$_1$ and \textit{base}$_2$, respectively.}
    \label{fig:base_pqcs}
\end{figure}
\begin{figure*}[t]
    \centering
    \includegraphics[width=0.65\linewidth, keepaspectratio]{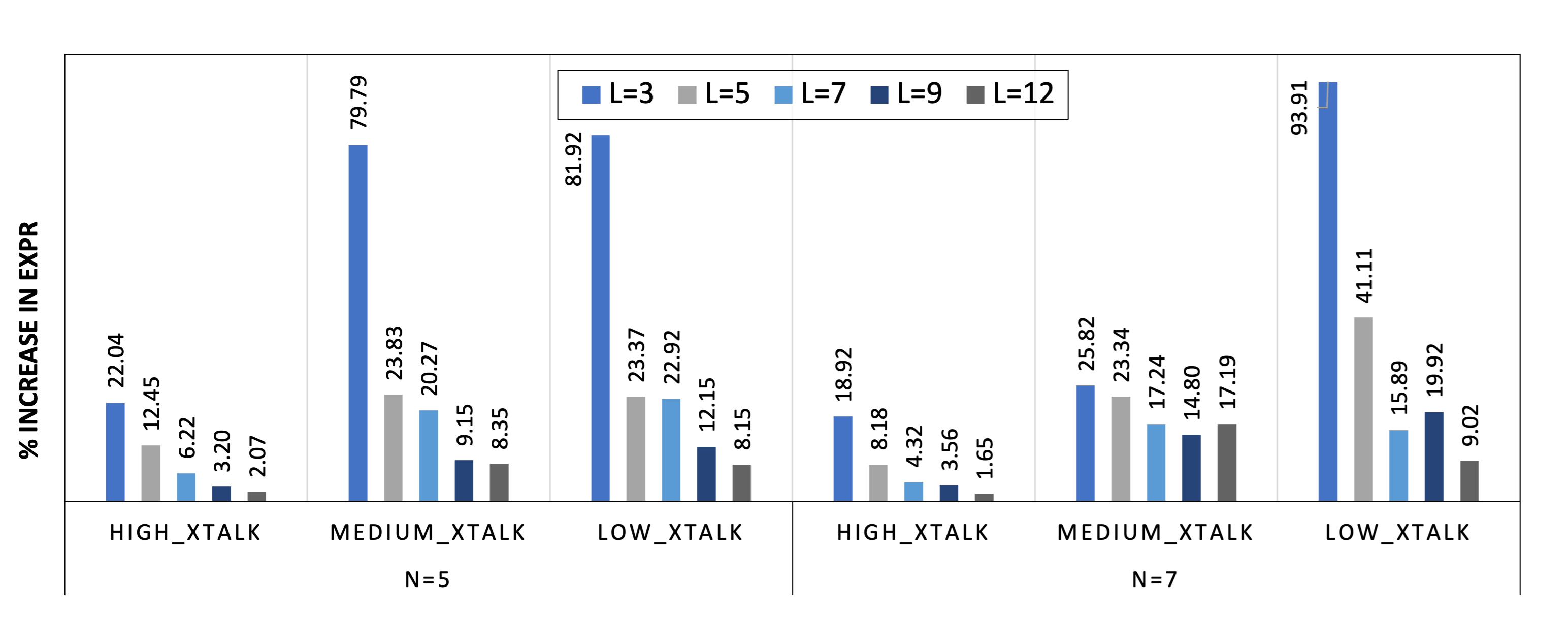}
    \caption{The effect of adding \textit{base}$_1$ layers on expressibility. The bars show the \% increase in expressibility for $5$- and $7$-qubit \textit{Xtalk\_pqcs} with $(m=2)$ over the same configurations with $(m=0)$ across different numbers of layers.}
    \label{fig:expr_base_vs_no_base}
\end{figure*}
Fig.~\ref{fig:base_pqcs} shows two base configurations we use in our evaluations (Section~\ref{sec:eval}). Fig.~\ref{fig:base_pqcs}(b) shows the single-layer configuration for \textit{base}$_1$, which contains the base entanglement layer we used in the previous step (before approximation) while Fig.~\ref{fig:base_pqcs}(c) shows the single-layer configuration for \textit{base}$_2$, which has the approximated entanglement layer (before applying XtalkSched).

With the sub-layers obtained from the scheduler, we can construct our \textit{Xtalk\_pqcs} as shown in Fig.~\ref{fig:clo_pqc}. The first $m$ layers of the PQC are from the base configuration. As \textit{Xtalk\_pqcs} can possibly have dispersed connectivity across a large number of layers, we add this option to help the optimized PQCs achieve better expressibility and entanglement. Our experimental analysis revealed that parameters such as expressibility saturate within $3$-$5$ layers. Therefore, we add up to $5$ base layers to our Xtalk circuit to make it more expressive. The rest of the circuit is constructed by alternating between single-qubit rotation layers and $R$ sub-layers obtained by the XtalkScheduler for $L-m$ times, where $L$ is the total number of layers.

\section{Results and Evaluation}
\label{sec:eval}
In this section, we first analyze different PQC configurations for expressibility, trainability, and entanglement. Next, we evaluate other circuit parameters such as duration, depth, and gate counts. Finally, we assess the PQCs' VQE performance on real hardware for two quantum chemistry benchmarks.

We evaluate five PQC configurations: \textit{base}$_1$, \textit{base}$_2$, \textit{high\_Xtalk}, \textit{medium\_Xtalk}, and \textit{low\_Xtalk}. Fig.~\ref{fig:base_pqcs} shows the configuration for \textit{base} PQCs. The \textit{Xtalk} circuits \{\textit{high\_Xtalk}, \textit{medium\_Xtalk}, \textit{low\_Xtalk}\} follow the configuration shown in Fig.~\ref{fig:clo_pqc}. The sub-layers are obtained by applying different levels of crosstalk-mitigation, with \textit{high\_Xtalk}, for example, corresponding to the lowest level of mitigation (Fig.~\ref{fig:xtalk_approx}(c)) and so on.
\subsection{Expressibility, Trainability, and Entanglement}
\subsubsection{Expressibility}
We first evaluate the effect of adding base layers to our \textit{Xtalk\_pqcs} on their expressibility. As mentioned in Section~\ref{sec:clo_method}, the sparsely connected sub-layers used in \textit{Xtalk\_pqcs} can possibly produce less expressive ansatze. An easy solution to this would be to add a few layers from the more expressive base configuration. 

Fig.~\ref{fig:expr_base_vs_no_base} shows the \% increase in expressibility due to adding two base$_1$ layers ($m=2$) for different $5$- and $7$-qubit \textit{Xtalk\_pqcs}. We pick a value of $2$ as our empirical analysis reveals that expressibility for various ansatz nears saturation at a number of layers in the range of $3$ to $5$. We see the highest increase in expressibility for shallower PQCs utilizing the sparsely connected \textit{Medium} and \textit{Low\_Xtalk} sub-layers ($79.7$ and $93.91\%$, respectively). We also see that the $7$-qubit \textit{Medium\_Xtalk} PQC with ($m=0$) is less susceptible to the addition of base layers than its $5$-qubit counterpart. This is an outcome of \textit{XtalkSchedule}'s performance with different circuit sizes. The $5$-qubit Medium and \textit{Low\_Xtalk} have very similar sub-layers' structure. On the other hand, the $7$-qubit PQCs give the scheduler more freedom to create different levels of crosstalk sensitivity. This led to the $7$-qubit \textit{Medium\_Xtalk} with ($m=0$) having more comparable expressibility to its ($m=2$) version and a lower increase. The relative difference drops as we increase the number of layers, which is expected as the expressibility of PQCs with ($m=0$) gradually increases. On average, the addition of $2$ base layers increases expressibility by $8.26$, $23.98$, and $32.84\%$ for \textit{High}, \textit{Medium}, and \textit{Low\_Xtalk} configurations.

\begin{figure}[!t]
    \centering
    \includegraphics[width=0.6\linewidth, keepaspectratio]{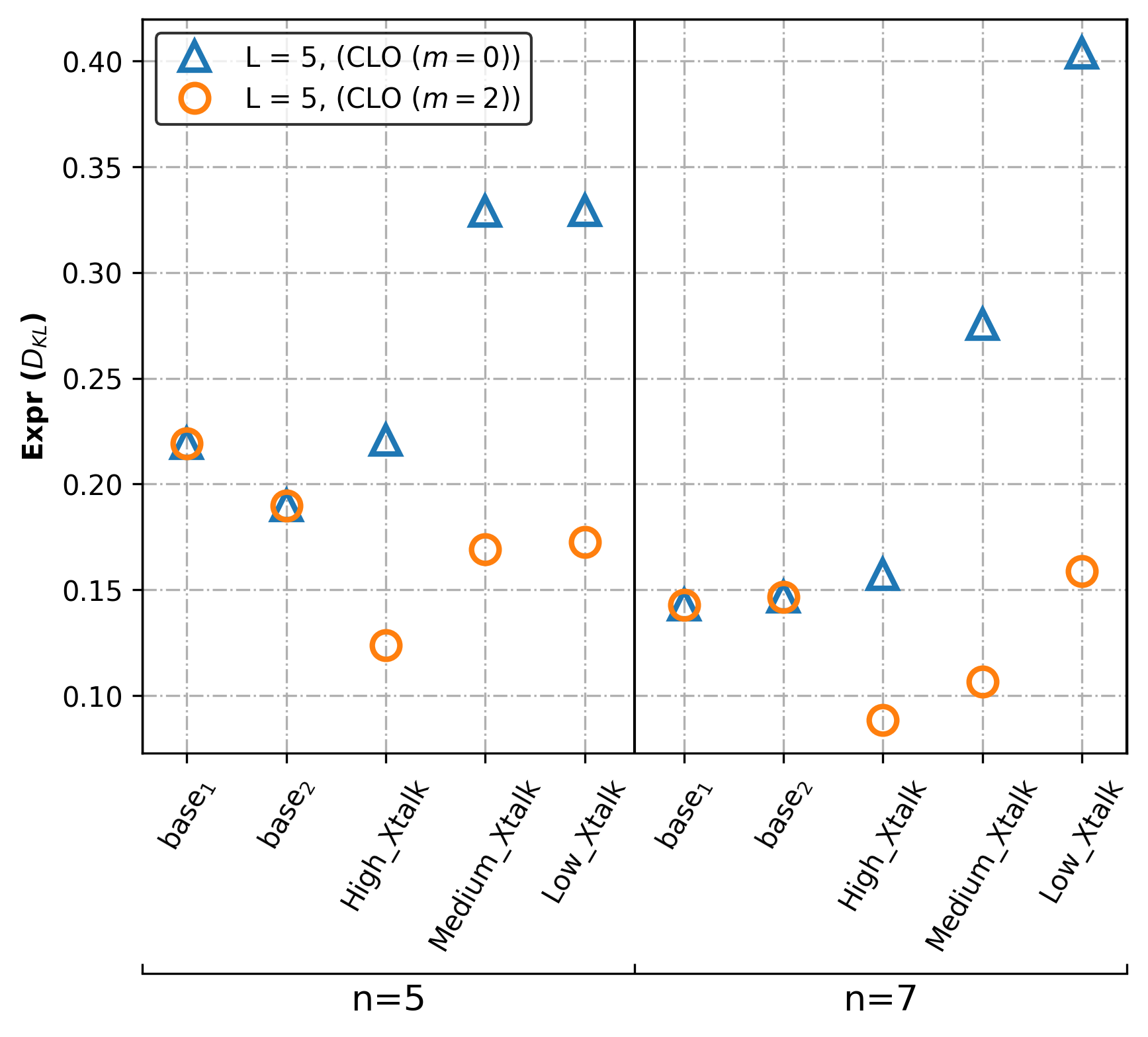}
    \caption{The expressibility of the different PQC configurations. The triangle and circular markers show the expressibility for \textit{Xtalk\_pqcs} with $(m=0)$ and $(m=2)$, resprectively.}
    \label{fig:clo_expr}
\end{figure}

Fig.~\ref{fig:clo_expr} shows the expressibility of the five PQC configurations used in our evaluations. We see that \textit{Xtalk\_pqcs} with ($m=0$) achieve similar (or better) expressibility to the base configurations. However, as demonstrated in prior research~\cite{holmes_expr, vqe_noise}, more expressibility does not directly lead to better performance. In fact, it can sometimes worsen a PQC's trainability~\cite{holmes_expr} for deep configurations. Thus, we leave identifying the number of $m$ layers leading to optimal expressibility as future work.
\begin{figure*}[!t]
    \centering
    \includegraphics[width=0.7\linewidth, keepaspectratio]{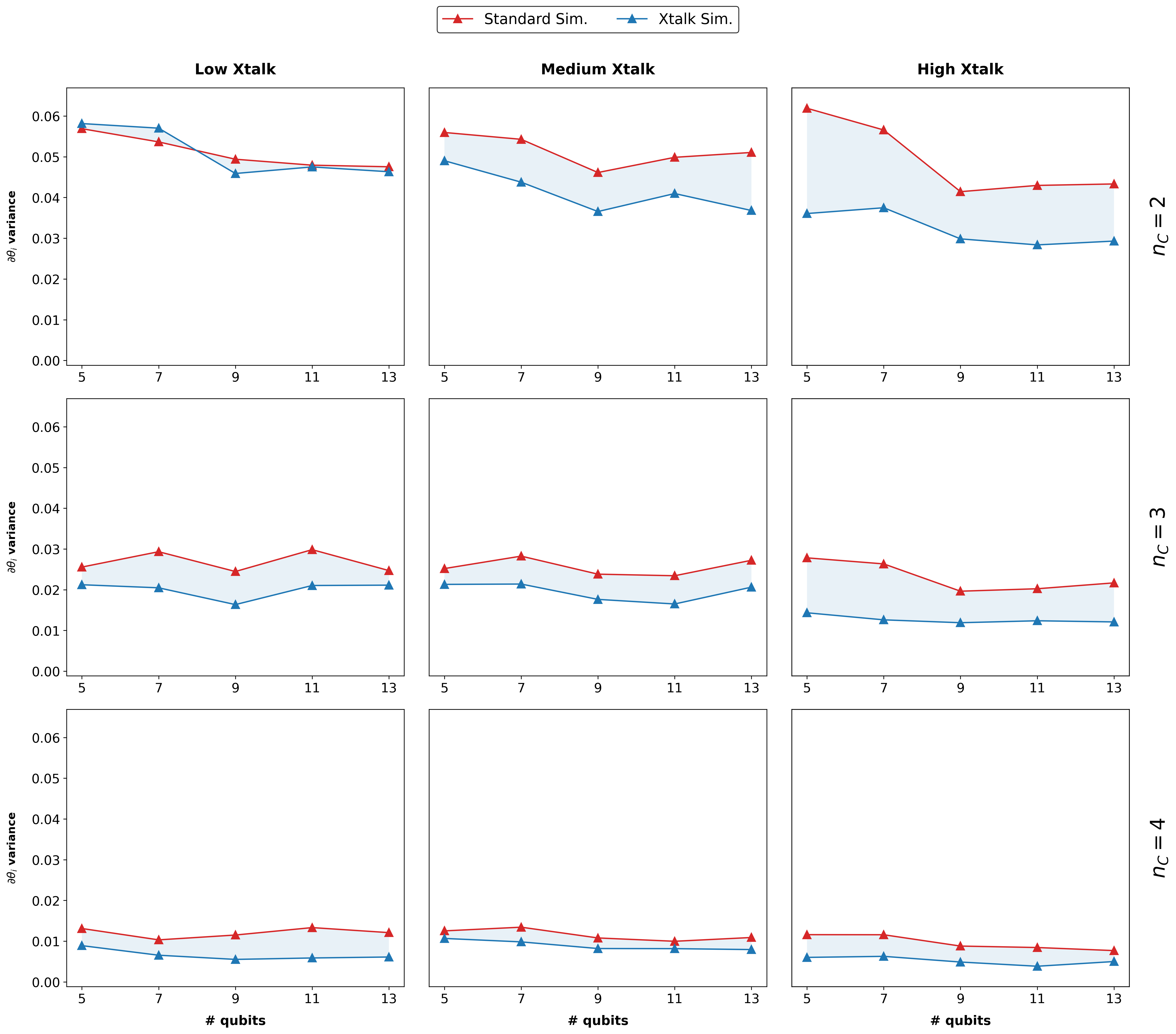}
    \caption{The change in variance of the partial cost function derivative for \textit{High}, \textit{Medium\_Xtalk}, and \textit{Low\_Xtalk} at different local cost function sizes ($n_C$), for shallow configurations ($L=\log_2(N)$), where $N$ is the number of PQC qubits. The figure also demonstrates the effect of crosstalk on trainability through experimenting with two types of simulation: \textit{Standard} QASM simulation and \textit{Xtalk-enabled} QASM simulation as described in Section~\ref{sec:expr_res}.}
    \label{fig:clo_trainability}
\end{figure*}
\subsubsection{Trainability}\label{sec:expr_res}
We conduct a cost-function-based trainability analysis~\cite{cerezo_bp} for the ground state preparation problem
\begin{equation}
\label{eqn:g_state}
C_G = 1 - p_{\ket{0}^{\otimes N}}
\end{equation}
where $N$ is the total number of qubits, and $p_{\ket{0}^{\otimes N}}$ is the probability of measuring the $\ket{00...0}_N$ state. For the local cost function, we only consider the probability of a subset of qubits
\begin{equation}
C_L = 1 - p_{\ket{0}^{\otimes N_C}}
\end{equation}
where $N_C$ is the number of cost-function qubits.

Qiskit's QASM simulator, which we use for this analysis, is currently limited to simulating independent gate errors. This poses a challenge for analyzing the impact of crosstalk on trainability. To address this issue, we modify the simulation to enable accounting for conditional error rates. This can be cheaply done by keeping a map of ``EPC\_multipliers" using the values indicated by Fig.~\ref{fig:xtalk_data}(b) color map. Next, for each DAG layer containing a parallel set of CNOTs, the multipliers map can be used to adjust their error rates. For example, consider a case when CNOT$_{0,1}$ is executing in parallel with CNOT$_{2,3}$ and CNOT$_{4,7}$ for which it has the multipliers 1.217 and 1.006, respectively. Then, its error rate for this particular instance will be its independent $EPC \times 1.217 \times 1.006$.

We analyze the trainability for \textit{High}, \textit{Medium}, and \textit{Low\_Xtalk} at different local cost function sizes ($n_C$), as shown in Fig.~\ref{fig:clo_trainability}. We see that crosstalk does indeed affect trainability, with the variance of the partial gradients decreasing by the increased crosstalk noise in the circuit. This suggests that barren plateaus can also be ``Crosstalk-induced", in addition to previous findings in~\cite{noise_bp} that only considered local Pauli noise. This observation is only available through the Xtalk-enabled simulation (blue lines in Fig.~\ref{fig:clo_trainability}), as the standard QASM simulation shows little to no variation between the different PQCs. Additionally, we observe that the effect of crosstalk decreases as we increase the cost function qubits ($n_C > 2$), as indicated by the shrinking size of the shaded regions.
\begin{figure*}[t]
\centering
\subfloat[\label{fig:entropy_fixed}]
  {\includegraphics[width=.4\textwidth]{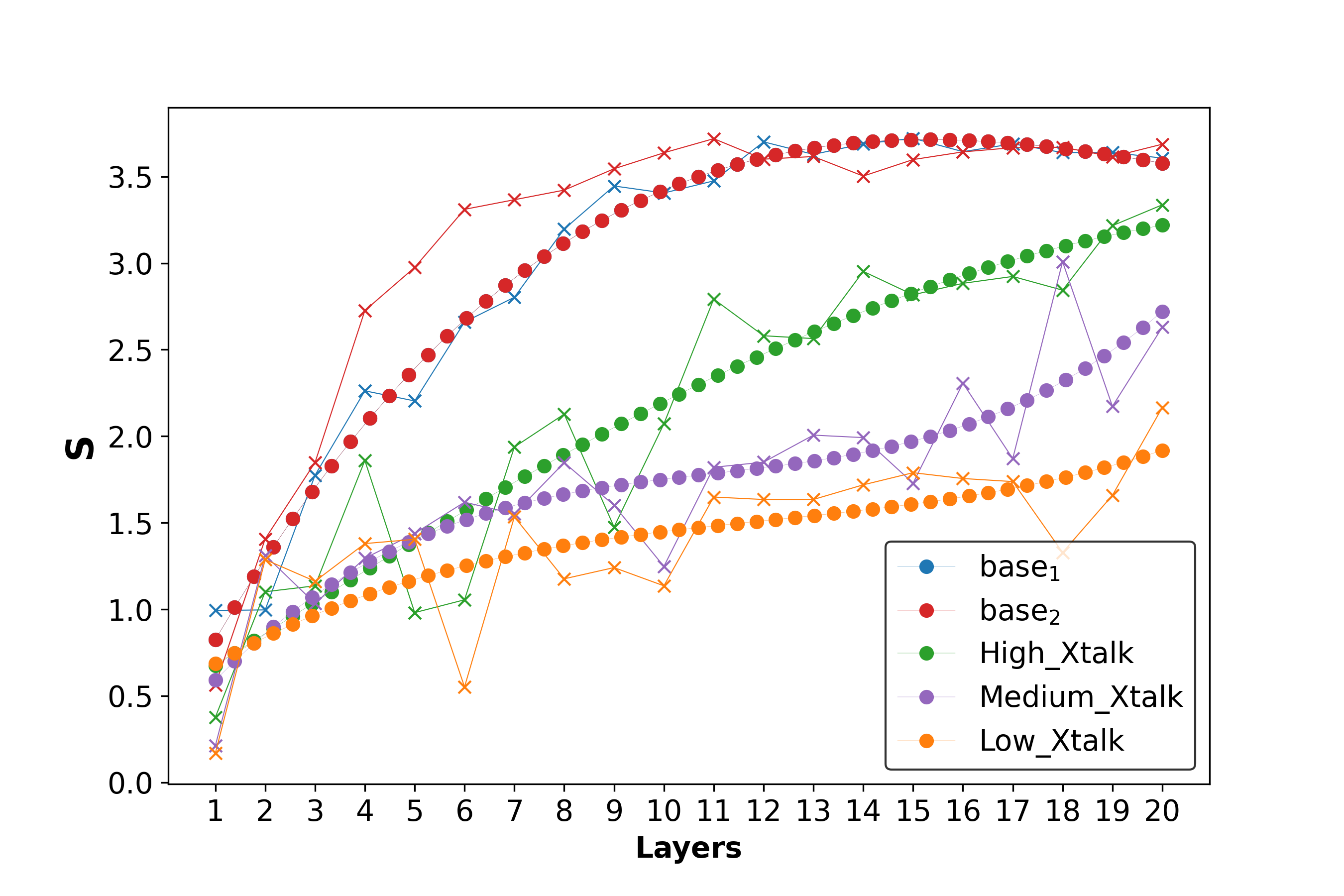}}\hspace{20pt}
\subfloat[\label{fig:entropy_inc}]
  {\includegraphics[width=.4\textwidth]{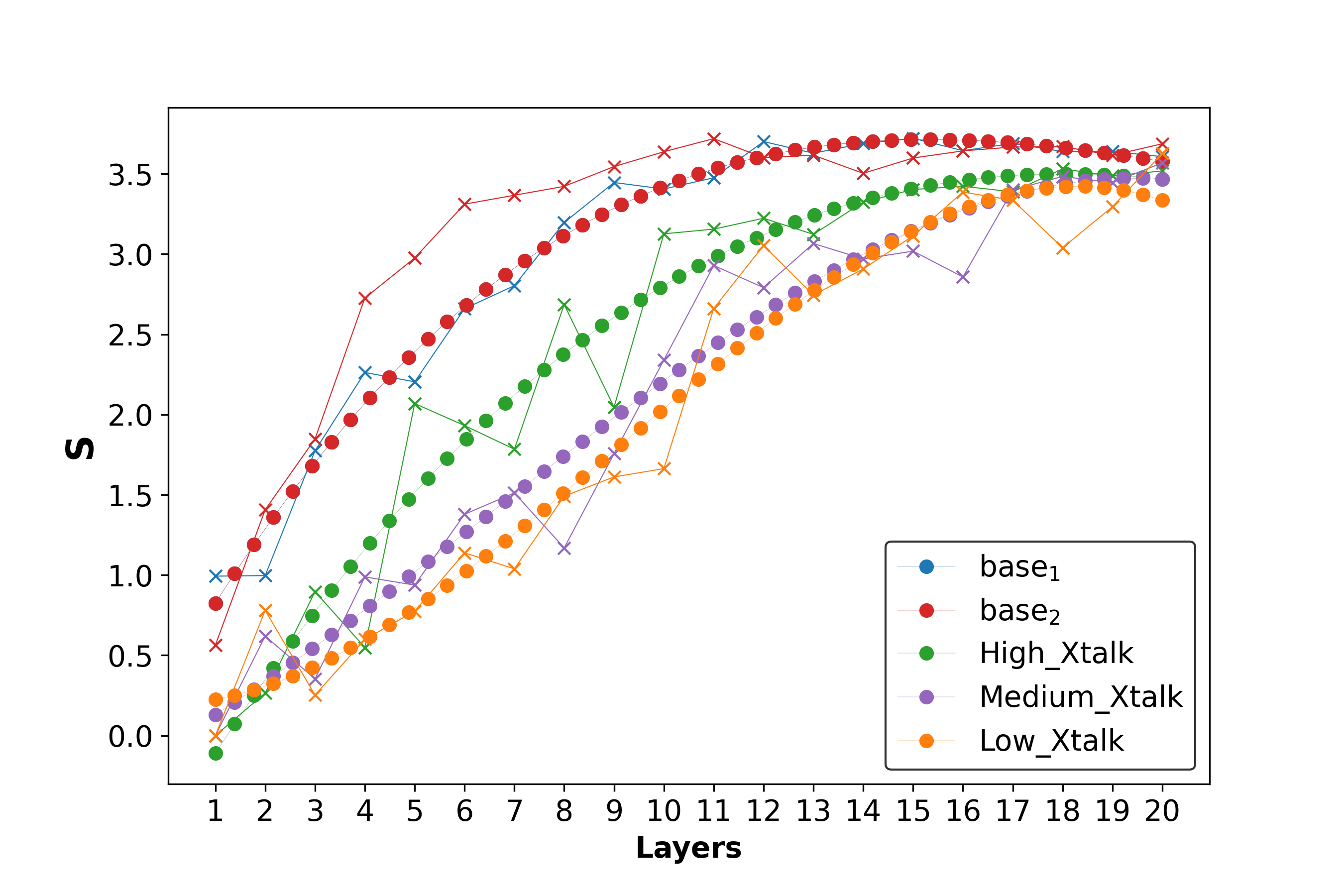}}\hfill
\caption{The trend lines of entanglement entropy $S$ for the five PQC configurations vs. number of layers. \textbf{(a)} Shows the entropy's trend with a fixed number of \textit{base} layers for the \textit{Xtalk\_pqcs} while \textbf{(b)} shows the trend with a number of \textit{base} layers that equals $1/3$ of total layers.}
\label{fig:entropy}
\end{figure*}
\subsubsection{Entanglement}
Fig.~\ref{fig:entropy_fixed} shows the trend of entanglement entropy for five PQC configurations with $9$-qubits and a $4$-$5$ partition. As expected, the base configurations create more entanglement compared to the Xtalk approach. Similar to expressibility, the entanglement trend is inversely proportional to the sparsity of the \textit{Xtalk\_pqc}'s sub-layers, with \textit{High\_Xtalk}'s trend line approaching base as depth increases. As previous work \cite{how_much_ent, ent_extra_1, ent_extra_2, ent_extra_3} states, different applications might require different levels of entanglement. A simple method to account for this entanglement loss in our \textit{Xtalk\_pqcs}, if an application necessitates it, is to increase $m$. Fig.~\ref{fig:entropy_inc} shows the trend in entanglement entropy for \textit{Xtalk\_pqcs} with ($m=\frac{1}{3} L$); entanglement quickly approaches the levels achieved by the base configurations.

\subsection{Experimental Setup}
We conducted our experiments on \textit{ibmq\_guadalupe}, a $16$-qubit backend available through IBM Quantum Service with average $T_1$ and $T_2$ times of $102.67$~$\mu$s and $108.06$~$\mu$s respectively, and an average CNOT error rate of \mbox{$1.013\mathrm{e}$-$2$} during the time of writing this paper. Note that these values fluctuate and are monitored through daily calibrations available through Qiskit. We utilize Qiskit Runtime~\cite{qiskit_runtime}, a programming model that allows for faster execution of quantum workloads on the cloud, to run our algorithm benchmarks.
\begin{figure*}[t]
    \centering
    \includegraphics[width=0.8\linewidth, keepaspectratio]{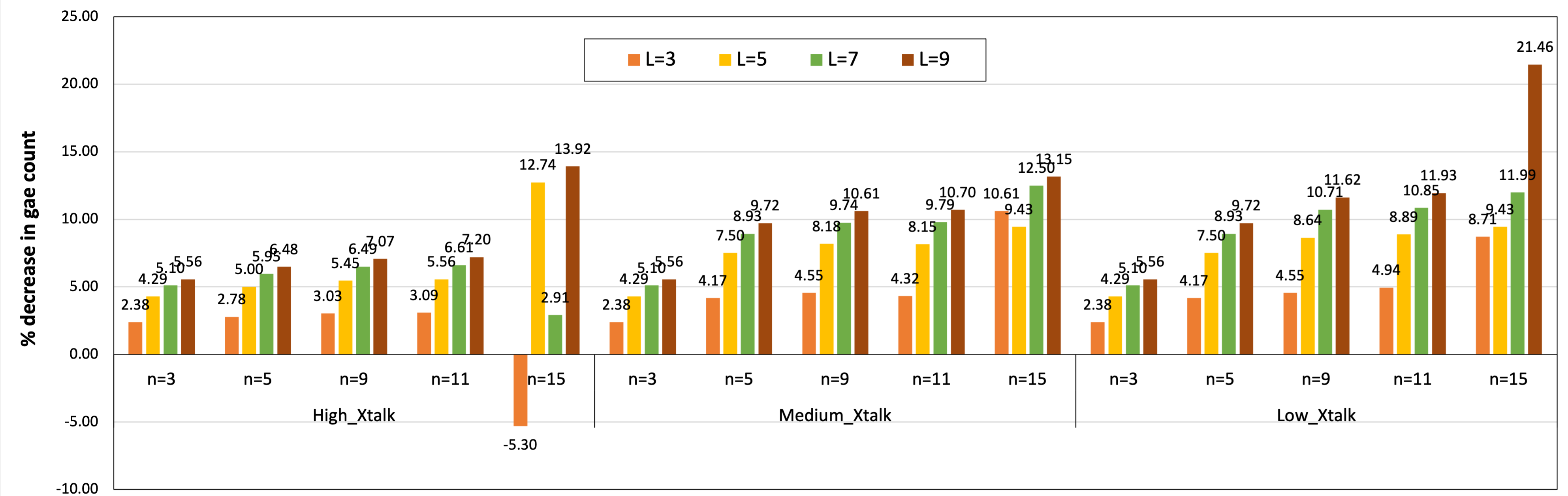}
    \caption{Decrease (\%) of total gate count over \textit{base} for different\textit{ Xtalk\_pqcs} with ($m=2$).}
    \label{fig:gate_counts}
\end{figure*}
\begin{figure}[t]
    \centering
    \includegraphics[width=0.6\linewidth, keepaspectratio]{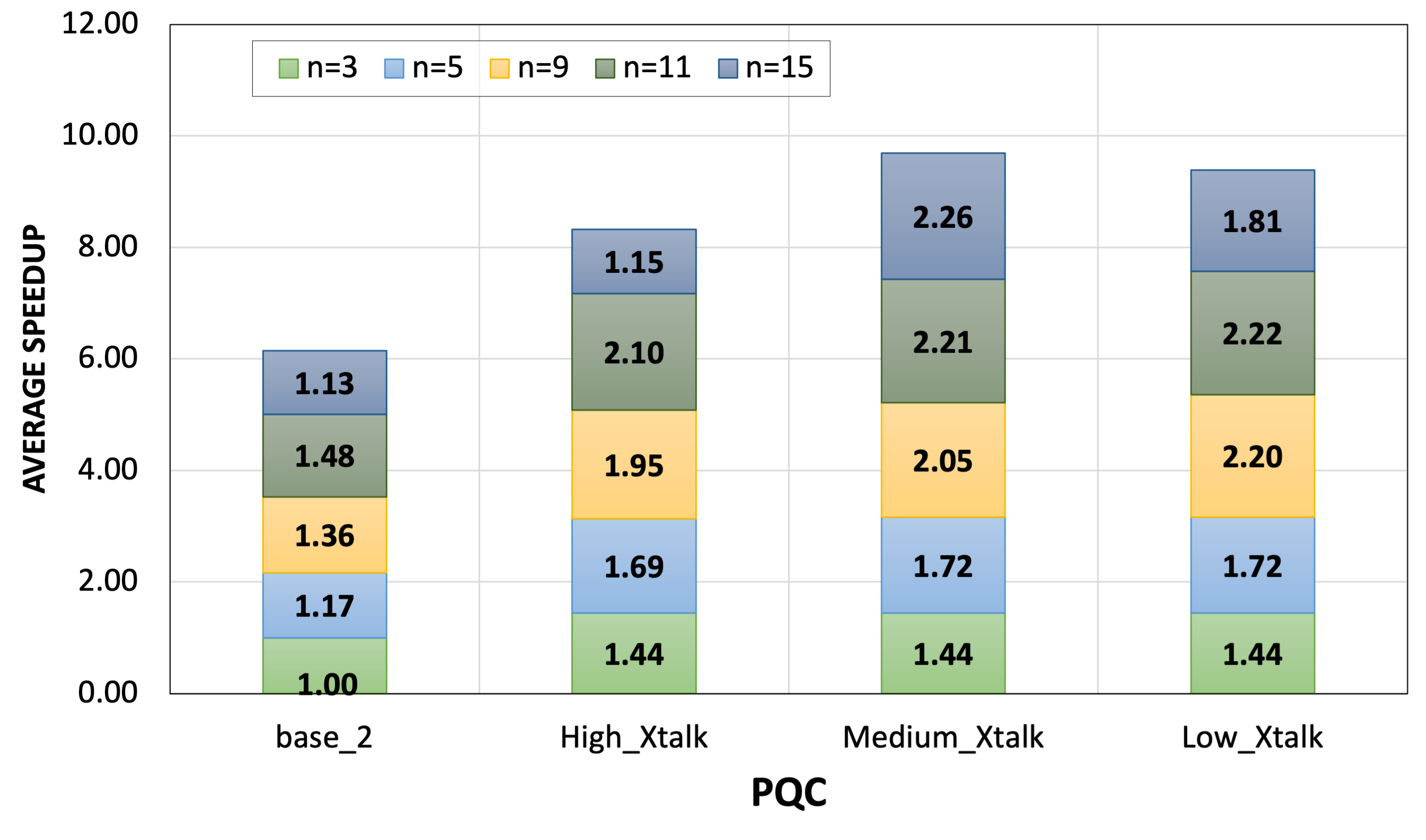}
    \caption{Average speed-up compared to \textit{base}$_1$ for \textit{base}$_2$ and \textit{Xtalk\_pqcs} with ($m=2$) at different circuit sizes.}
    \label{fig:durations}
\end{figure}
\subsection{Circuit Parameters}
We record the total gate count and duration of each PQC with different configurations. We compile each configuration with three Qiskit optimization levels and report the best gate count and duration for each.

Fig.~\ref{fig:gate_counts} shows the percentage of total gate count reduction for each \textit{Xtalk\_pqc} (with $m=2$) compared to \textit{base}. It is important to note that both base configurations (\textit{base}$_1$ and \textit{base}$_2$) have very similar gate counts as they both share the same pre-compiled number of gates and their suitability to $1$D mapping (mapping to a line of qubits). Additionally, our Xtalk-based approach does not change the number of single-qubit gates; thus, the gate reductions observed are ultimately two-qubit gate reductions. We see an expected outcome that all \textit{Xtalk\_pqcs} reduce the number of two-qubit gates, as each layer (after the first two layers) has a number of operations less than \textit{base}. Therefore, the percentage of reduction grows with increasing the number of layers for each configuration, as indicated by the figure. Overall, \textit{Xtalk\_pqcs} have an average gate count reduction of $5.7$, $7.97$, and $8.57\%$ for \textit{High}, \textit{Medium}, and \textit{Low\_Xtalk} configurations, respectively, with up to $21.46\%$ for the $15$-qubit \textit{Low\_Xtalk} PQC. This specific higher-than-average value of $21.46\%$ is not directly attributed to our Xtalk approach. We argue that it is due to the compiler's utilization of the reduced number of gates in its optimization approach. 

Fig.~\ref{fig:durations} shows the average speedups achieved by \textit{base}$_2$, \textit{High}, \textit{Medium}, and \textit{Low\_Xtalk} compared to \textit{base}$_1$. We first note the difference between \textit{base}$_1$ and \textit{base}$_2$. Unlike total gate count, \textit{base}$_2$ PQCs have lower depths as a result of the approximation to an alternating structure. Therefore, we see that \textit{base}$_2$ speeds up the execution time with an average of $1.23\times$ at different PQC sizes and up to $1.48\times$. \textit{Xtalk\_pqcs}, on the other hand, observe higher average speedups due to their lower depths compared to both base PQCs. The average speedups are $1.66\times$, $1.94\times$, and $1.88\times$ for \textit{High}, \textit{Medium}, and \textit{Low\_Xtalk}, respectively, with up to $2.93\times$ and $2.86\times$ reported for the latter two configurations at ($L=3$, $n=15$). We also observe that the rate of speedup drops for the $15$-qubit \textit{base}$_2$ and \textit{High\_Xtalk} ($1.13\times$ and $1.15\times$ respectively). This is due to the mapping of the circuits on the $16$-qubit \textit{ibmq\_guadalupe}. As the circuit size nears the backend's total number of qubits, the compiler will be unable to perform $1$D mapping. Therefore, it is more likely to add SWAP gates to satisfy all operations in \textit{base}$_2$ and \textit{High} compared to \textit{Medium} and \textit{Low\_Xtalk}, which will be easier to map due to their lower number of operations and hence, possible easier mappings to the limited connectivity.
\subsection{Algorithm Performance}
\label{sec:clo_results}
We evaluate the performance of the PQCs for finding the ground state energy of the H$_2$ and LiH molecules through VQE, which corresponds to finding the minimum eigenvalue of Hermitian matrices characterizing these molecules. We configured our experiments and PQCs as follows. We ran all our benchmarks on \textit{ibmq\_guadalupe} accessed through IBM Cloud. We picked the Simultaneous Perturbation Stochastic Approximation (SPSA)~\cite{spsa} as the classical optimizer, with a maximum of $100$ iterations. The number of PQC layers $L$ was set to $5$ for both benchmarks, with ($m=2$) for \textit{Xtalk\_pqcs}. We obtained both Hamiltonians using Bravyi-Kitaev (BK) fermionic mapping technique~\cite{bk} with Active-Space reduction~\cite{Rossmannek_2021}, resulting in $4$- and $6$-qubit Hamiltonians for H$_2$ and LiH, respectively. We chose BK-based Hamiltonians over other mapping techniques (e.g., Jordan-Wigner or Parity~\cite{jw}) that result in lower-qubit Hamiltonians and, in return, might perform better on hardware~\cite{mond_hpqc}. The reason for this choice is that our three Xtalk variants (\textit{High}, \textit{Medium}, and \textit{Low}) are more distinguishable at larger PQC sizes.\footnote{The experiment's primary goal is to explore the differences between Xtalk variants, not to get the most chemically accurate result.}
\begin{figure*}[t]
\centering
\subfloat[\label{fig:clo_H2}]
  {\includegraphics[width=.35\textwidth]{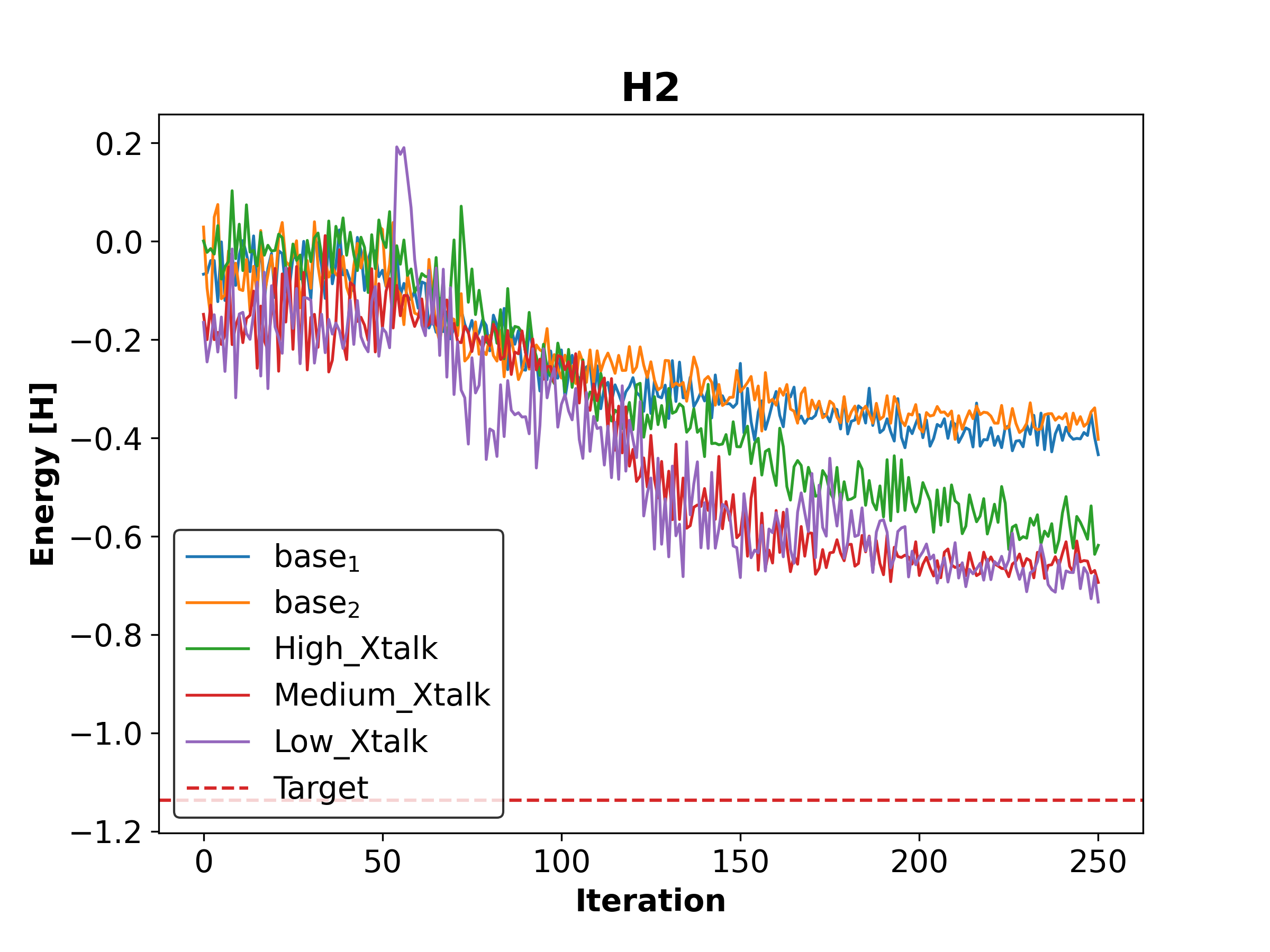}}\hspace{30pt}
\subfloat[\label{fig:clo_LiH}]
  {\includegraphics[width=.35\textwidth]{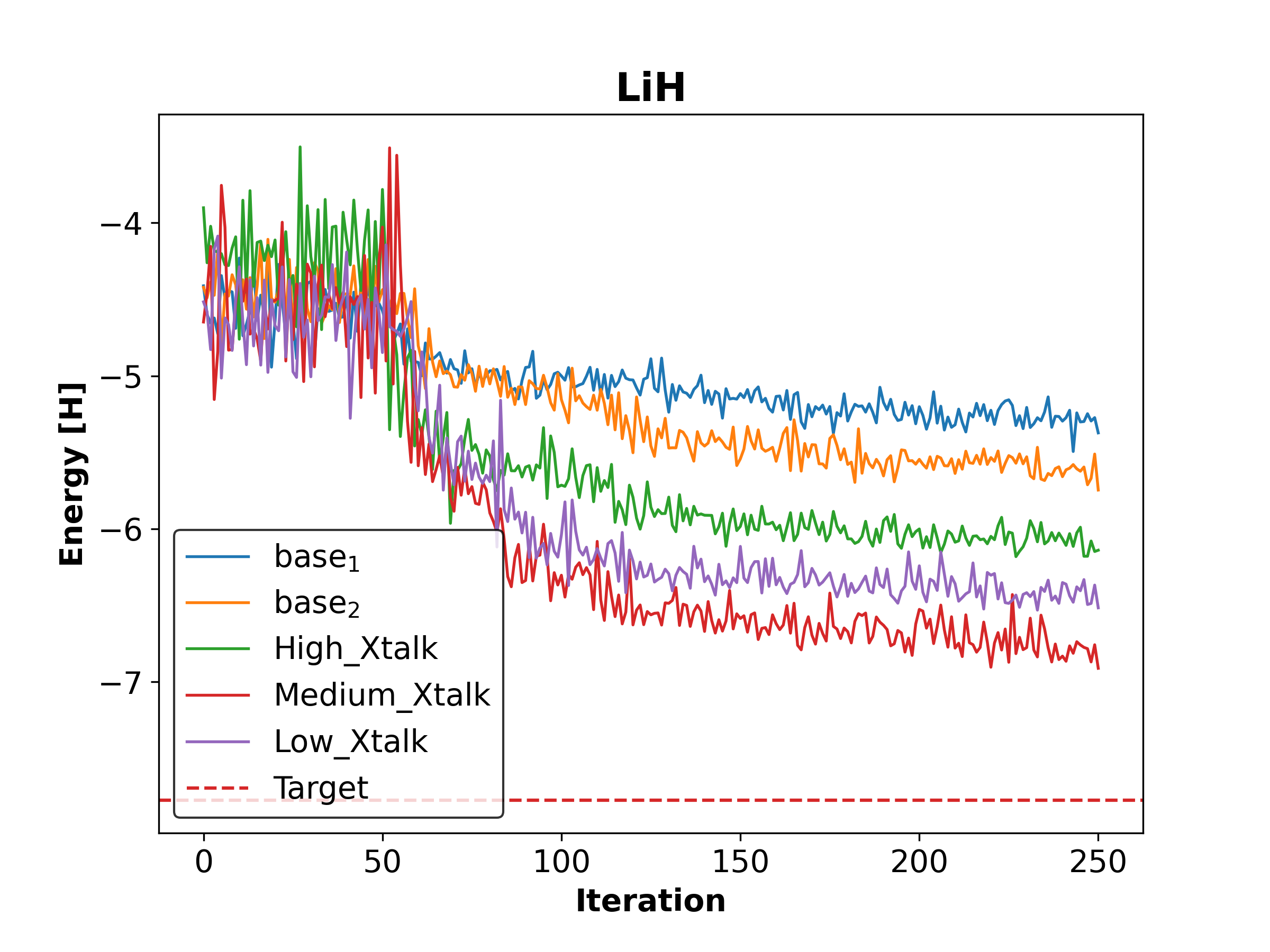}}\hfill
\caption{VQE results for H$_2$ and LiH molecules.}
\label{fig:clo_chem}
\end{figure*}
Figures.~\ref{fig:clo_H2} and~\ref{fig:clo_LiH} show VQE results for H$_2$ and LiH molecules respectively. Although all PQC configurations do not reach the target ground state energies, \textit{Xtalk\_pqcs} clearly outperform base configurations for both benchmarks. We make two observations from the figures. First, both figures confirm the advantage of our Xtalk approach and the effect of crosstalk mitigation on algorithm performance. Second, Fig.~\ref{fig:clo_LiH} shows that the best performing PQC is \textit{Medium\_Xtalk}. This suggests that the level of crosstalk-based approximation should be tailored to each application to achieve the best results, which we leave as future work.
\section{Related Work}
Crosstalk-based approaches for compilation and execution have been investigated in~\cite{murali_xtalk, ding_xtalk, niu}. Niu \textit{et al.}~\cite{niu} investigated parallel execution techniques in noisy quantum hardware by comparing the state-of-the-art methods and discussing their shortcomings with the impact of various aspects. Consequently, they proposed a Quantum Crosstalk-aware Parallel workload execution method (QuCP) with no crosstalk characterization overhead and additionally utilized their method to parallelize Zero Noise Extrapolation (ZNE) workloads and reduce their cost. Ding \textit{et al.}~\cite{ding_xtalk} introduced a systematic methodology for software mitigation of crosstalk due to the frequency crowding phenomenon. Their strategy allows for fixed coupler architectures to have matching levels of reliability to tunable coupler architectures, thus simplifying quantum machines' fabrication. While this work trades parallelism with higher gate fidelity when needed, it dramatically improves the resilience of tunable qubits in fixed-couple hardware.

Hardware-oriented compilation and approximation for VQAs have gained much attraction recently~\cite{qnas, patel_approx, gushu_isca}. Wang \textit{et al.}~\cite{qnas} proposed a noise-adaptive co-search framework for variational circuits and qubit mapping, which utilizes iterative pruning to remove redundant gates in the searched circuits. Their investigation suggested several routes for more theoretical and experimental exploration in variational quantum algorithms, with one route, the variational ansatz, being optimized to alleviate barren plateaus. Patel \textit{et al.}~\cite{patel_approx} proposed a method for reducing CNOT gate count by generating approximations for quantum circuits through partitioning for scalability. Their method reduces the circuit length with approximate synthesis while improving fidelity by running circuits representing key samples in the approximation space. The work proposed by Li \textit{et al.}~\cite{gushu_isca} leverages Pauli strings to identify program components and introduce optimizations at the algorithm, compiler, and hardware levels, for a family of chemistry problems.

\section{Conclusions}
In this paper, we examined a new approach to embedding a machine's characteristics in the VQA ansatze design. We developed a strategy to approximate PQCs to a more hardware-efficient version by utilizing the hardware's crosstalk characteristics. Our approach aims at creating a version of the ansatz that inherently mitigates crosstalk by utilizing crosstalk-based scheduling. The methodology can be used to create approximated PQCs with various levels of crosstalk mitigation.

Our analysis shows that crosstalk mitigation enhances the performance of VQE. We utilized a combination of hardware and algorithmic PQC analysis parameters to evaluate our Xtalk approach. Our results demonstrate that the Xtalk approach maintains similar expressibility to a pre-approximated \textit{base} and is more trainable for local a cost function, all while speeding up the execution by an average of $1.83\times$ (up to $2.93\times$) and reducing the total gate count by an average of $7.9\%$ (up to $21.46\%$). Moreover, our algorithm performance results show that \textit{Xtalk\_pqcs} clearly outperform \textit{base} for estimating the ground state of two chemical molecules using VQE. Furthermore, the results hint that, although \textit{Xtalk\_pqcs} generally perform better than \textit{base}, the level of crosstalk mitigation used to construct a \textit{Xtalk\_pqc} is not directly proportional to its algorithmic performance. Therefore, a method that closely ties the approximation degree to the application's performance should will be explored as future work.

\section{Acknowledgement}
M.I. would would like to thank the NSF QISE-NET Fellowship for funding through the grant DMR 17-47426, and the IBM Quantum Hub at NC State for access to \textit{ibmq\_guadalupe}.
\bibliographystyle{IEEEtran}
\bibliography{references}
\end{document}